\documentclass{article}\usepackage[]{graphicx}\usepackage[]{color}
\makeatletter
\def\maxwidth{ %
  \ifdim\Gin@nat@width>\linewidth
    \linewidth
  \else
    \Gin@nat@width
  \fi
}
\makeatother

\definecolor{fgcolor}{rgb}{0.345, 0.345, 0.345}

\usepackage{framed}
\makeatletter
 {\par\unskip\endMakeFramed%
 \at@end@of@kframe}
\makeatother

\definecolor{shadecolor}{rgb}{.97, .97, .97}
\definecolor{messagecolor}{rgb}{0, 0, 0}
\definecolor{warningcolor}{rgb}{1, 0, 1}
\definecolor{errorcolor}{rgb}{1, 0, 0}
\newenvironment{knitrout}{}{} 

\usepackage{alltt}

\usepackage{amsmath}
\usepackage{natbib}

\newtheorem{theorem}{Theorem}
\newtheorem{remark}{Remark}
\IfFileExists{upquote.sty}{\usepackage{upquote}}{}
\begin{document}

\title{Selective correlations - the conditional estimators}
\author{Yoav Benjamini\footnotemark[1] \footnotemark[2]  and Amit Meir\footnotemark[3]}

\maketitle

\footnotetext[1]{Department of Statistics and Operations Research, The Sackler Faculty of Exact Sciences
 Tel Aviv University}
\footnotetext[2]{The Sagol School of Neuroscience, Tel Aviv University}
\footnotetext[3]{Department of Statistics, University of Washington}

\begin{abstract}
The problem of Voodoo correlations is recognized in neuroimaging as the problem of estimating quantities of interest from the same data that was used to select them as interesting. In statistical terminology, the problem of inference following selection from the same data is that of \emph{selective inference}. Motivated by the unwelcome side-effects of the recommended remedy- splitting the data. A method for constructing confidence intervals based on the correct post-selection distribution of the observations  has been suggested recently. We utilize a similar approach in order to provide point estimates that account for a large part of the selection bias. We show via extensive simulations that the proposed estimator has favorable properties, namely, that it is likely to reduce estimation bias and the mean squared error compared to the direct estimator without sacrificing power to detect non-zero correlation as in the case of the data splitting approach. We show that both point estimates and confidence intervals are needed in order to get a full assessment of the uncertainty in the point estimates as both are integrated into the Confidence Calibration Plots proposed recently.  

The computation of the estimators is implemented in an accompanying software package.
\end{abstract}

\section{Introduction}
In the pursuit of brain regions that are highly correlated with behavioral measures (neural correlates), past practice has been to report correlations between the imaging measurements and behavioral attributes only in selected regions- usually these are detected using the same correlations to be reported.

\begin{knitrout}
\definecolor{shadecolor}{rgb}{0.969, 0.969, 0.969}\color{fgcolor}\begin{figure}[ht]

\includegraphics[width=\maxwidth]{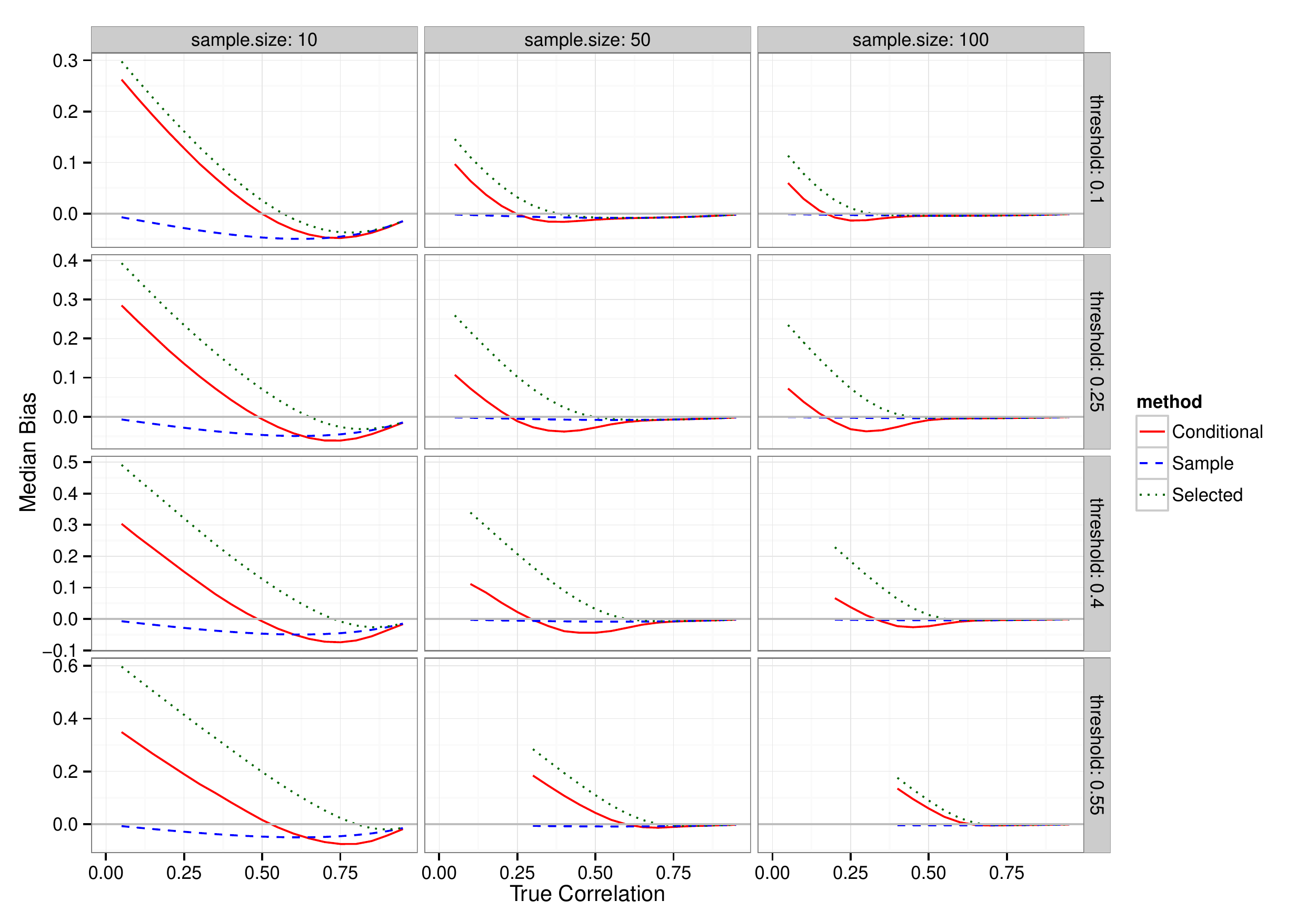} \caption[The median bias for the entire sample and the selected subset]{The median bias for the entire sample and the selected subset. The figure demonstrates that selection bias is present whenever a non-independent data-driven parameter selection has been perfromed. Biases are plotted for the correlation estimate on the entire data, the correlation estimate on the selected subset and for the conditional method proposed in this work. The true underlying correlations varies from 0.05 to .95 (x axis), and is the same for all observations. The number of subjects underlying each observed corrleation varies from 10 to 100 (in columns). Selection was perfromed according to a constant threshold.\label{fig:introduction}}
\end{figure}

\end{knitrout}

The implication of such unattended \emph{selective estimation} have been raised recently by two provocative papers \citet{vul2009puzzlingly} and \citet{button2013power}. The problem raised by \citet{vul2009puzzlingly} is essentially that reported correlations between imaging attributes are "puzzilingly high". These papers have been so influential that the title of the former paper- "Voodoo Correlations"- has become an unofficial term for selection bias. \cite{rosenblatt2014voodoo} give an overview of the paper and the ensuing discussion in the scientific literature.

\begin{knitrout}
\definecolor{shadecolor}{rgb}{0.969, 0.969, 0.969}\color{fgcolor}\begin{figure}[h]

\includegraphics[width=\maxwidth]{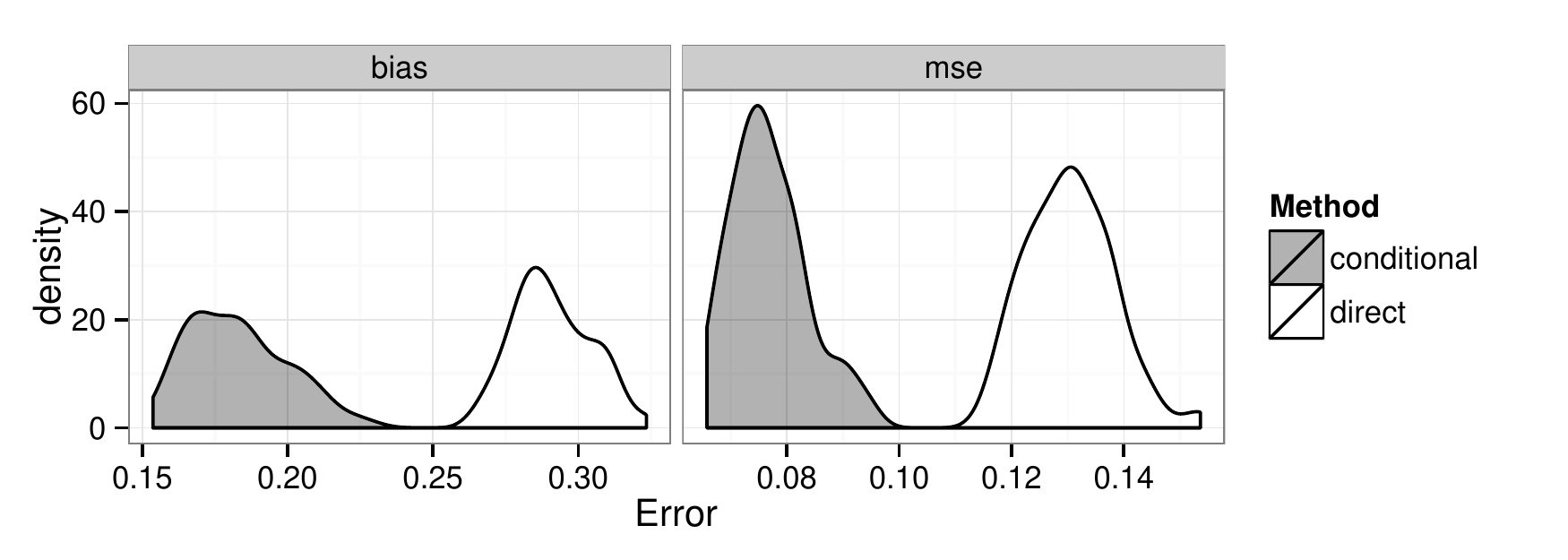} \caption[Comparison of the direct esimtates and the conditional estiamtes when applied to data similar in structure to fMRI data]{Comparison of the direct esimtates and the conditional estiamtes when applied to data similar in structure to fMRI data. These results are based on 200 artificial data sets genrated by the process described in the appendix. For each data set a seperate BH selection threshold was determined and the mean squared error (MSE) and bias were computed for the selected voxels. The plotted densities describe the distribution of the simulated MSE and bias\label{fig:ErrorDensPlot}}
\end{figure}

\end{knitrout}

We wish to point to the fact that estimation bias exists whenever non-independent selection occurs and can be quite considerable. Even large observed correlations, say $r=0.8$, can stem from non-existing ones merely due to noise. Bias occurs in the presence or in the absence of a true effect. Bias will be present even if flawless control of multiplicity is performed. In fact, the more conservative the multiplicity control, the higher the selection threshold so that only extreme values survive it. Finally, bias will occur in very large sample sizes,  though it obviously decreases with sample sizes: the larger the sample size, the smaller the standard errors of the estimated correlation. The lower the selection threshold, the milder the selection bias. The existence of bias in selective inference is presented in Figure \ref{fig:introduction}. In the absence of selection the median bias is nearly zero for all correlation values. However, when selection is present there is a significant upward bias for all but the highest true correlation values. The conditional estimation method proposed in this article is shown to reduce this bias.

Imposing independence by splitting the data was the recommended remedy in \citet{vul2009puzzlingly} and shared by almost all commentators (\citet{kriegeskorte2010everything};\citet{fiedler2011voodoo};\citet{poldrack2009independence};\citet{lazar2009discussion};\citet{lindquist2009correlations};\citet{nichols2009commentary};\citet{yarkoni2009big}), While remedying bias, splitting the data introduces variance effects making it an unattractive method  when dealing with small samples. Splitting the data both reduces the power to select and increases the variance of the estimators for the selected observations.

An alternative approach has been proposed by \citet{rosenblatt2014voodoo}. Instead of splitting the data, they propose to construct confidence intervals for the selected correlations in a way such that the  \emph{False Coverage Rate} (FCR) is controlled. In the aforementioned paper, two methods are advocated for controlling the FCR. The first, FCR-adjusted confidence intervals utilizes an approach first advocated by \citet{benjamini2005false}. The second method, Conditional Quasi Conventional Confidence Intervals (CQC CI), constructs confidence intervals based on the probability distribution of the observed correlation conditioned on the observed correlation being above the selection threshold. This method was first proposed by \citet{weinstein2013selection}. \citeauthor{rosenblatt2014voodoo} present theory and a simulation study to demonstrate that the proposed methods indeed control the FCR.

While \citeauthor{rosenblatt2014voodoo}'s method provides us with a valid confidence intervals for the selected correlations, thereby quantifying the additional uncertainty induced by the selection,  they do not address the problem of providing improved point estimates. With the aim of performing point estimation for the selected variables, we follow the approach of \citeauthor{rosenblatt2014voodoo} and base our inference on the distribution of the estimator conditioned on it being above some threshold. To the best of our knowledge, this approach was first introduced by \citet{hedges1985statistical} and by \citet{iyengar1988selection} in the context of meta-analysis who used it to adjust estimators for being significant at $.05$ level. 

In an extensive simulation study, we show that the proposed method is preferable to the data splitting approach as it has more power to identify active brain regions and provides point estimates that are less variable. In section \ref{sub:fmribased} we compare the conditional estimates to the direct correlation estimates used for selection by applying both methods to artificial data which has similar structure to real fMRI data. The main results of the simulation are presented in Figure \ref{fig:ErrorDensPlot}, it is shown that the the proposed method has both lower bias and lower overall mean squared error than the direct estimates.

The layout of the article is as follows, in section \ref{sec:method} we present the proposed conditional estimation method for performing post selection inference. In section \ref{sec:simstudy} we conduct a simulation study with the aim of investigating the behavior of the proposed method when applied to data with structure similar to that of fMRI data. In section \ref{sec:fMRI} we apply our method to data from the well known study by \citet{tom2007neural}. Finally, section \ref{sec:conclusion} concludes.

\section{Conditional Maximum Likelihood}\label{sec:method}
\begin{knitrout}
\definecolor{shadecolor}{rgb}{0.969, 0.969, 0.969}\color{fgcolor}\begin{figure}[h]

{\centering \includegraphics[width=\maxwidth]{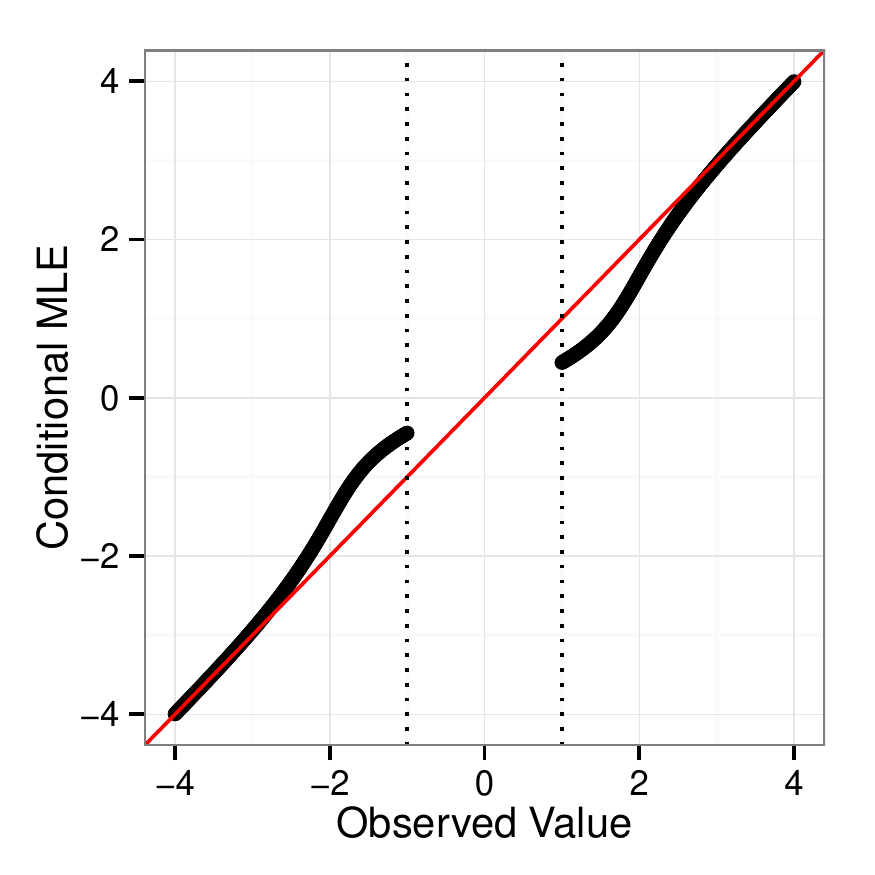} 

}

\caption[Demonstration of the behaviour of the conditional estimator for $Y\sim\mathcal{N}(0,1)$ with threshld $c=1$]{Demonstration of the behaviour of the conditional estimator for $Y\sim\mathcal{N}(0,1)$ with threshld $c=1$. The conditional estimator acts as a compromise between soft and hard thresholding procedures. When the observed value gets close to the threshold the estimator of $\theta$ is shrunk towards zero while for observed values that are far away from the threshold hardly any shrinkage is done. Ofcourse no value is available for $|Y|<c$ as these are of no interest.\label{fig:shrinkage}}
\end{figure}

\end{knitrout}

In this section we present the conditional estimation method for estimating selected parameters. Let $Y\sim f_{Y;\theta}$ be a random variable that possesses a CDF $F_Y(y;\theta):=F_{Y;\theta}(y)=P_\theta(Y<y)$. In addition, assume that we are only interested in the value of the parameter $\theta$ if $|Y|$ is big enough, say bigger than $c>0$. Alternatively, we only observe $X\stackrel{d}{=}(Y||Y|>c)$. This conditional distribution depends on $\theta$. Let $Q_c(\theta)=P(|Y|\geq c)=1+F_{Y,\theta}(-c)-F_{Y,\theta}(c)$, then the probability density of $X$ is given by
\begin{equation}
f_{X,\theta}(x)=\frac{f_{Y,\theta}(x)}{Q_c(\theta)}I\{|x|\geq c\}.
\end{equation}
In other words, the density of the observed random variable $X$ is zero on $(-c,c)$ and proportional to that of $Y$ elsewhere. Note that the conditioning may alter the role of the parameter $\theta$; For example, if we have $Y\sim\mathcal{N}(\theta,1)$, while $\theta$ is a mere location parameter for $Y$, it reflects both the location and shape of $X$. In order to obtain an estimate for $\theta$, we can use any standard estimation approach such as Maximum Likelihood or Moment Estimation. Here, we obtain point estimates for $\theta$ through Maximum Likelihood.                                 

In this article, we are mainly interested in the case of $Y\sim \mathcal{N}(\theta,\sigma^{2})$ with $\sigma^{2}$ known. In this particular case finding the MLE of $f_{X,\theta}$ is straight forward due to the fact that the likelihood function is continuous and unimodal as a function of $\theta$. The behavior of the conditional estimator is demonstrated in Figure \ref{fig:shrinkage} for the case $Y\sim N(\theta,1)$ where $Y$ is observed if $|Y|>1$. The conditional estimator acts as a compromise between a soft and hard thresholding procedures (see for example, \cite{donoho1995adapting}). When the observed value of $Y$ is close to the threshold, for example, $Y=1.05$, the MLE is shrunk all the way down to $0.47$. On the other hand, if the observed value of $Y$ is far away from the threshold, for example, $Y=3.5$, then the estimator is $\hat{\theta}=3.48$. Note that as $Y$ is normally distributed, then the MLE is equivalent to the moment estimator since $X$ belongs to the exponential family.

We wish to apply the conditional estimation method to estimating selected correlations. This can be done by applying Fisher's transform to the observed correlation. Let $\rho_i$ be the true correlation of the $i$th observation and $r_i$ be the observed correlation coefficient, then 
\begin{equation}
Y_i=\frac{1}{2}\log\left(\frac{1+r}{1-r}\right)\approx\mathcal{N}\left(\frac{1}{2}\log\frac{1+\rho}{1-\rho},\frac{1}{n-3}\right).
\end{equation}
Applying Fisher's transform to the observed correlations enables us to easily compute the conditional estimator for 
\begin{equation}
\theta:=\frac{1}{2}\log\left(\frac{1+\rho}{1-\rho}\right)
\end{equation}
We then apply the inverse Fisher transform to the conditional estimator $\hat{\theta}_c$ to obtain a conditional estimator for $\rho$ 
\begin{equation}
\hat{\rho}_c=\frac{e^{2\hat{\theta}_c}-1}{e^{2\hat{\theta}_c}+1}
\end{equation}
where we are assured that $\hat{\rho}_c$ is indeed the conditional estimator for $\rho$ by the monotone functional invariance of the MLE. 

\begin{remark}
So far we considered estimation when using a fixed predetermined threshold. The threshold can be determined by Random Field Theory that ignores the cluster size or by Bonferroni- both make use of a predetermined threshold. However, in other cases we may be interested in performing selection in such a way as to control quantities such as the FDR by using the Benjamini Hochberg (BH) procedure or any other FDR controlling procedure. An issue which arises in the case of selection rules such as BH, is that the threshold is data dependent and not constant. In a simulation study, \cite{rosenblatt2014voodoo} showed that CQC CIs indeed control the FCR when conditioned on the effective threshold determined by the data. This practice can also be justified using a result by \citet{storey2004strong}: they show that under general conditions the BH threshold converges to a constant as the number of hypothesis grow. Therefore, conditioning on the BH threshold is consistent for conditioning on the fixed value to which the BH threshold converges. This subject if further discussed in the appendix A. 
\end{remark}

\section{Simulation Study}\label{sec:simstudy}
In this section we present the properties of the conditional estimation method when used to estimate correlations. We compare our method with two other methods- the direct estimator and the 50-50 split method. The direct estimator is the standard correlation value that was used to perform selection. The 50-50 split is a method that seeks to eliminate the selection bias by splitting the data into equal sized subsets and performing the selection on one subset and the estimation of the selected correlations on the other.

In section \ref{sub:sampselect} we compare the behavior of the estimators in a simple setting and demonstrate how the accuracy of the estimators change as we switch from measuring our error on the entire sample to measuring our error over the selected subset. In section \ref{sub:mix} we evaluate the estimators under a more realistic setting where the data is dependent and generated from a mixture model and the threshold is determined by the data. In section \ref{sub:fmribased} we take a step further in trying to emulate real fMRI data by   applying the estimators to a data set produced based on real fMRI data.

\subsection{Measuring Quality Over the Selected vs. Entire Sample}\label{sub:sampselect}

\begin{knitrout}
\definecolor{shadecolor}{rgb}{0.969, 0.969, 0.969}\color{fgcolor}\begin{figure}[ht]

\includegraphics[width=\maxwidth]{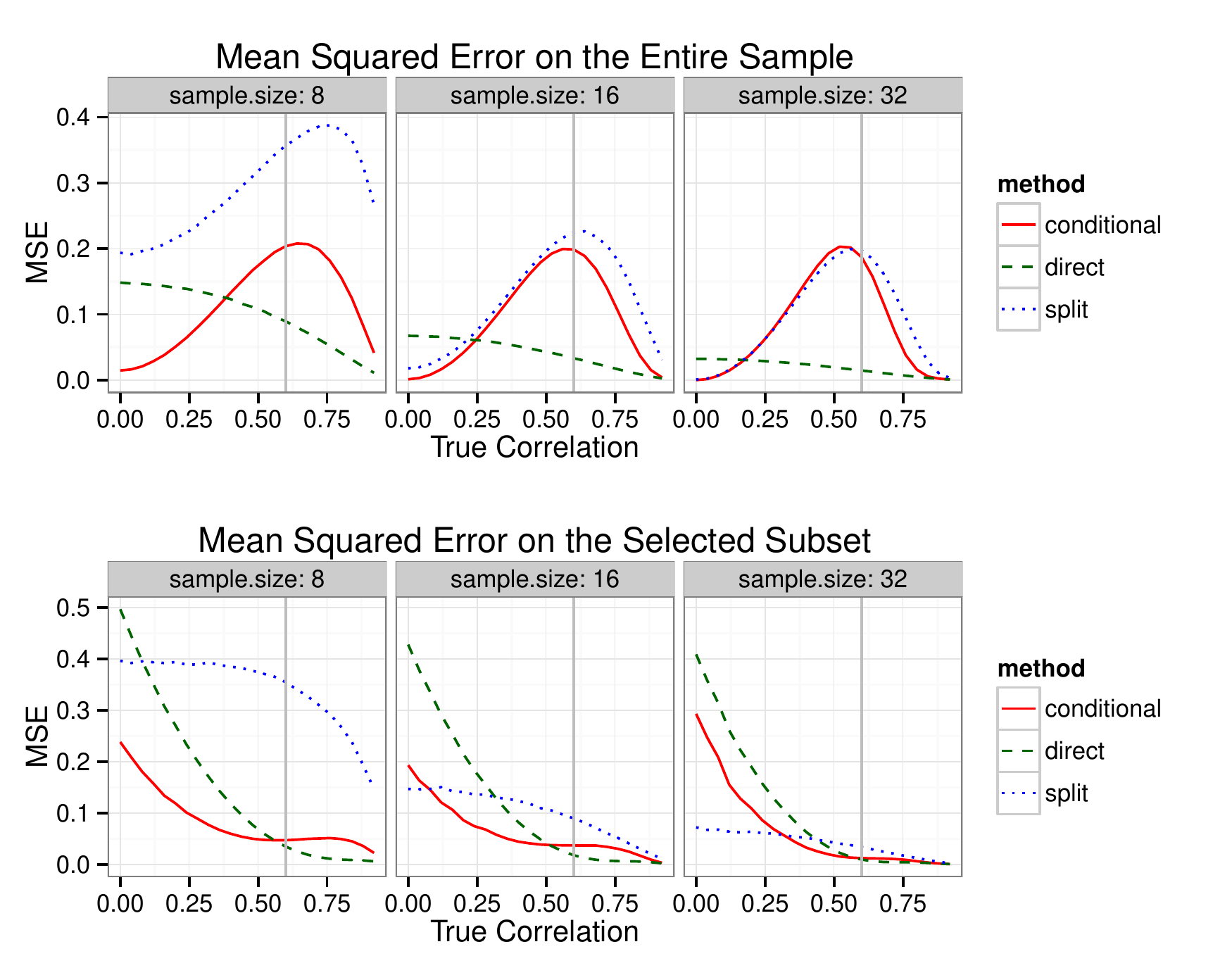} \caption[The mean squared error of the different method on the entire sample and selected observations]{The mean squared error of the different method on the entire sample and selected observations. For the split and conditional methods the estimators for observations not passing the selection threshold of 0.6 received a value of zero.\label{fig:risk.plot}}
\end{figure}

\end{knitrout}

In this section we evaluate the estimators under the simple setting of independent correlations, with the goal of demonstrating how their relative performance differs when we applied to the whole data or just the selected subset. In Figure \ref{fig:risk.plot} we plot the the risk of the conditional, split and direct estimators as a function of the true correlation value for different sample sizes on the entire sample and on the observations which pass the threshold of $0.6$. 

When we assess the behavior of the estimators on the entire sample, it is clear that the direct estimator performs best for higher correlation values whereas the conditional estimator performs best for lower correlation values where it benefits from the thresholding of higher observed correlation values. For the smaller sample sizes the conditional estimator performs better than the 50-50 split because it has more power to recognize large correlations and has smaller variance. 

On the subset of observations that pass the threshold the direct correlation estimates suffer from a considerable amount of bias, especially when the true correlation is below the threshold. When the true correlation is high,  the direct estimator outperforms the selective estimation methods. For the small and medium size samples the conditional method clearly outperforms the 50-50 split method while on the largest sample size the 50-50 split method performs the best for small correlation values and is surpassed by the conditional method for high correlation values. An important fact demonstrated in Figure \ref{fig:risk.plot} is that no method dominates the others in all circumstances (even if we are only interested in the error on the selected correlations).

\subsection{A More Realistic Mixture Model}\label{sub:mix}

\begin{knitrout}
\definecolor{shadecolor}{rgb}{0.969, 0.969, 0.969}\color{fgcolor}\begin{figure}[p]

\includegraphics[width=\maxwidth]{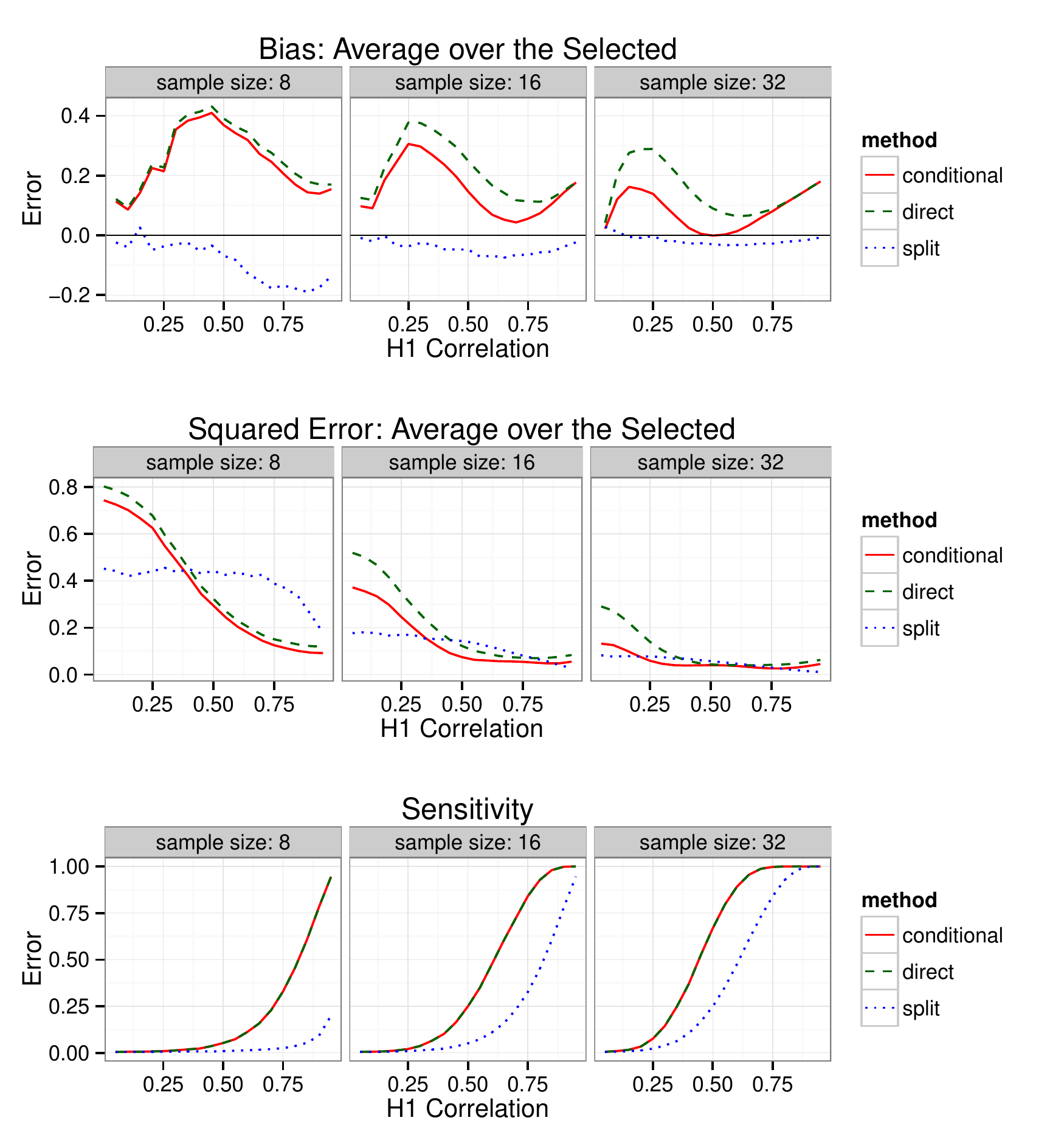} \caption[Simulated average MSE of the proposed estimation methods under dependence for selected voxels]{Simulated average MSE of the proposed estimation methods under dependence for selected voxels. Voxel selection was performed using the BH selection rule ($\alpha=0.1$). Observations constructed as a sum of a signal field and a smooth Gaussian noise field. The H1 corrleation displays the underlaying correlation of the signal varying from 0.05 to 0.95. Each line represents a different number of subjects varying through 8,16 and 32. The propensity of non-null correlation is 0.2. Simulated 'brains' consist of $10\times 10\times 10$ voxels. If no voxels were selected then no MSE was reported.\label{fig:mixture.plot}}
\end{figure}

\end{knitrout}

In this section, for simulation purposes we assume that the sampling distribution of Pearson's correlation, after Gaussinization with a Fisher transformation, behaves like a smooth Gaussian stationary random field. Observation were hence constructed as a sum of signal field and a smooth Gaussian noise field where the signal was produced from a mixture distribution. The propensity of non-zero correlations was set to $0.2$, similar to the one found by a mixture of Gaussians fit to the data collected by \citet{tom2007neural}. The results of the simulation are quite insensitive to the propensity of non-zero correlations.

The results of the simulation is presented in Figure \ref{fig:mixture.plot}.  For all sample sizes and true correlation values the conditional method has lower bias and MSE than the direct estimator. This can be be explained by the conditional method having lower risk for small correlation values that make up most of the underlying correlations in the simulated data. As could be expected, the 50-50 split method exhibits almost no bias and therefore outperforms the other methods with respect to bias. On the other hand, when we examine the average MSE, the 50-50 split method is outperformed by the competing methods for all but the lowest correlation values. This is mostly due to the higher variance of the estimates. 

While the 50-50 split method being bias free is an attractive property, the bias and MSE on the selected subset of correlations do not tell the entire story. As stated by some of the commentators on \citet{vul2009puzzlingly}'s paper, the main interest of researchers conducting fMRI studies is detecting brain regions whose activity levels are correlated with the measure of interest (see for example \citet{lieberman2009correlations}). As demonstrated in the last row of Figure \ref{fig:mixture.plot}, the 50-50 split method is significantly under powered when it comes detecting non-zero correlations. For example, when we have a sample size of $32$ and the value of the true non-null correlation is $0.5$ then the 50-50 split method detects $25\%$ of the non-null voxels where using the entire sample for selection detects about $75\%$ of the non-null voxels. 

Notice that while in Figure \ref{fig:risk.plot} the difference between the MSE of direct estimator and the MSE of the conditional estimator decreases with the sample size, in Figure \ref{fig:mixture.plot} the difference between the MSEs increases with the sample size. This can be explained by the fact that in Figure \ref{fig:mixture.plot} the threshold is data dependent and its value decreases with the number of observations. For a constant threshold, as the number of observations grow and the variance of the observations decreases the observations values for which the conditional estimator performs shrinkage decreases. When the threshold decreases with the variance the conditional estimator performs shrinkage for a larger portion of the observed values as lower observed values are possible. 

\subsection{fMRI Based Data}\label{sub:fmribased}
While in the previous subsection we assumed simple mixture distributions for the signal and GRF distribution for the noise, in this subsection we take a step further in trying to emulate the structure of fMRI data. We base our data generation process on fMRI data collected by \citet{tom2007neural}. 

Again, we assume that after Gaussianization, the observed fMRI data is the sum of a signal, the distribution of which is unknown to us, and a GRF noise. In order to approximate fMRI data we extract a signal from the raw data and add GRF noise with average variance of $(n-3)^{-1}$. More details regarding the data generation process are given in appendix \ref{app:fmri}. The resulting data are presented in Figure \ref{fig:fMRIdemonstration} together with the original data. The artificial data is quite similar in structure to the real data albeit being a bit smoother. We report the result of applying the proposed method to 200 such data sets.

The first set of results presented in Figure \ref{fig:ErrorDensPlot} are the distribution of the MSE and mean bias obtained from the 200 replications of the experiment. It is clear that not accounting for the selection process results in a significantly higher bias and mse. The mode of the bias and the mse of the direct estimates are approximately 0.28 and 0.13 respectively while the modes of the bias and the mse of the conditional method are approximately 0.175 and 0.08 respectively.

\begin{knitrout}
\definecolor{shadecolor}{rgb}{0.969, 0.969, 0.969}\color{fgcolor}\begin{figure}[t]

\includegraphics[width=\maxwidth]{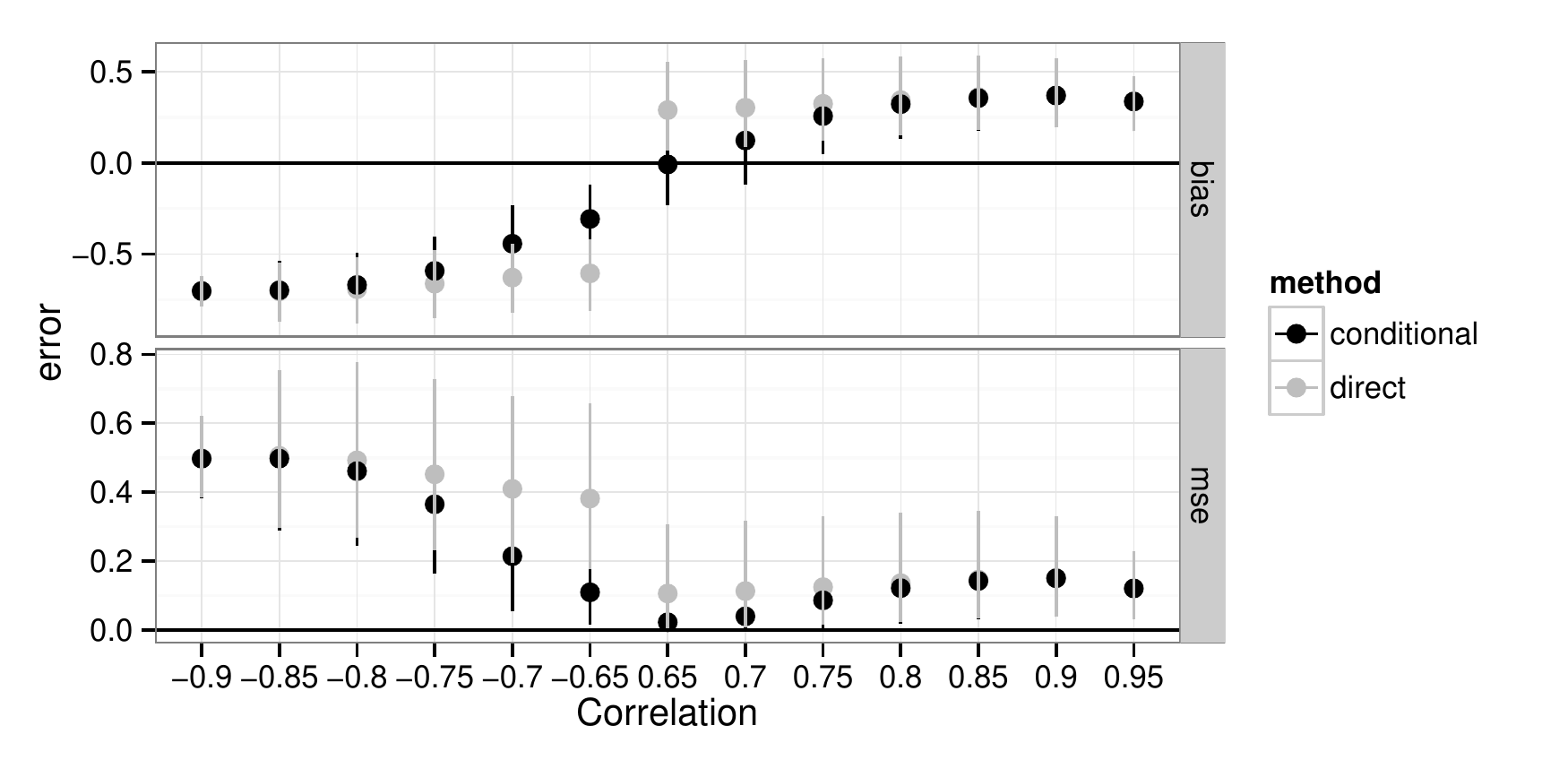} \caption[Simulated averge mse and bias as a function of the observed correlations for the conditional method and the direct estimator for selected voxels]{Simulated averge mse and bias as a function of the observed correlations for the conditional method and the direct estimator for selected voxels. The mse and bias were computed as in Figure \ref{fig:ErrorDensPlot}. The point estimates for the mse and bias are plotted with vertical bars depicting the 0.05 and 0.95 percentiles of the distribution.\label{fig:byobserved}}
\end{figure}

\end{knitrout}

In Figure \ref{fig:byobserved} we plot the bias and mse as a function of the observed correlation with the $0.05$ and $0.95$ percentiles of the distribution. First, it appears that the absolute upward bias of the direct estimator is roughly constant $0.3$ when the observed value is positive and in the neighborhood of $-0.7$ when the observed value is negative. For the conditional estimator, the bias becomes worse as the observed value is further away from zero. When the observed value is close to the threshold $\approx 0.65$, the estimation bias is zero on average. But as the observed values go further away from the threshold the estimation bias becomes worse and for observed absolute values of $0.85$ or higher the conditional estimator is nearly identical to the direct estimator and therefore the mse and bias are also identical.   

\section{Application to fMRI Data}\label{sec:fMRI}
We can now approach social-neuroscience studies such as the ones discussed by \citet{vul2009puzzlingly}. In particular, the study performed by \citet{tom2007neural}. In this high profile study which was revisited in replies to \citet{vul2009puzzlingly}, the authors attempted to localize brain regions associated with the individuals' loss-aversion. This was done by correlating the behavioral loss-aversion index of 16 subjects with a neural loss-aversion index at each voxel. The data was organized, documented and kindly made available via \emph{openFMRI} initiative at https://openfmri.org/dataset/ds000005.

In that study high correlations were reported in 8 selected brain regions. These regions were selected using hypothesis tests on a robust version of this same correlation. \citet{poldrack2009independence} later confirmed that these findings were indeed the result of unaccounted selective estimation and when controlling for the selection using a data split, an average upward selection bias of about 0.3 was discovered. 

We present the results of applying the conditional MLE method to this data via the "Confidence Calibration Plot" first introduced by \citet{rosenblatt2014voodoo} in Figure \ref{fig:CCPplot}. This is a parametric map which's legend is augmented by CQC CIs presented in the aforementioned paper and in our case, also augmented by the conditional MLEs.

\begin{knitrout}
\definecolor{shadecolor}{rgb}{0.969, 0.969, 0.969}\color{fgcolor}\begin{figure}[pt]

\includegraphics[width=\maxwidth]{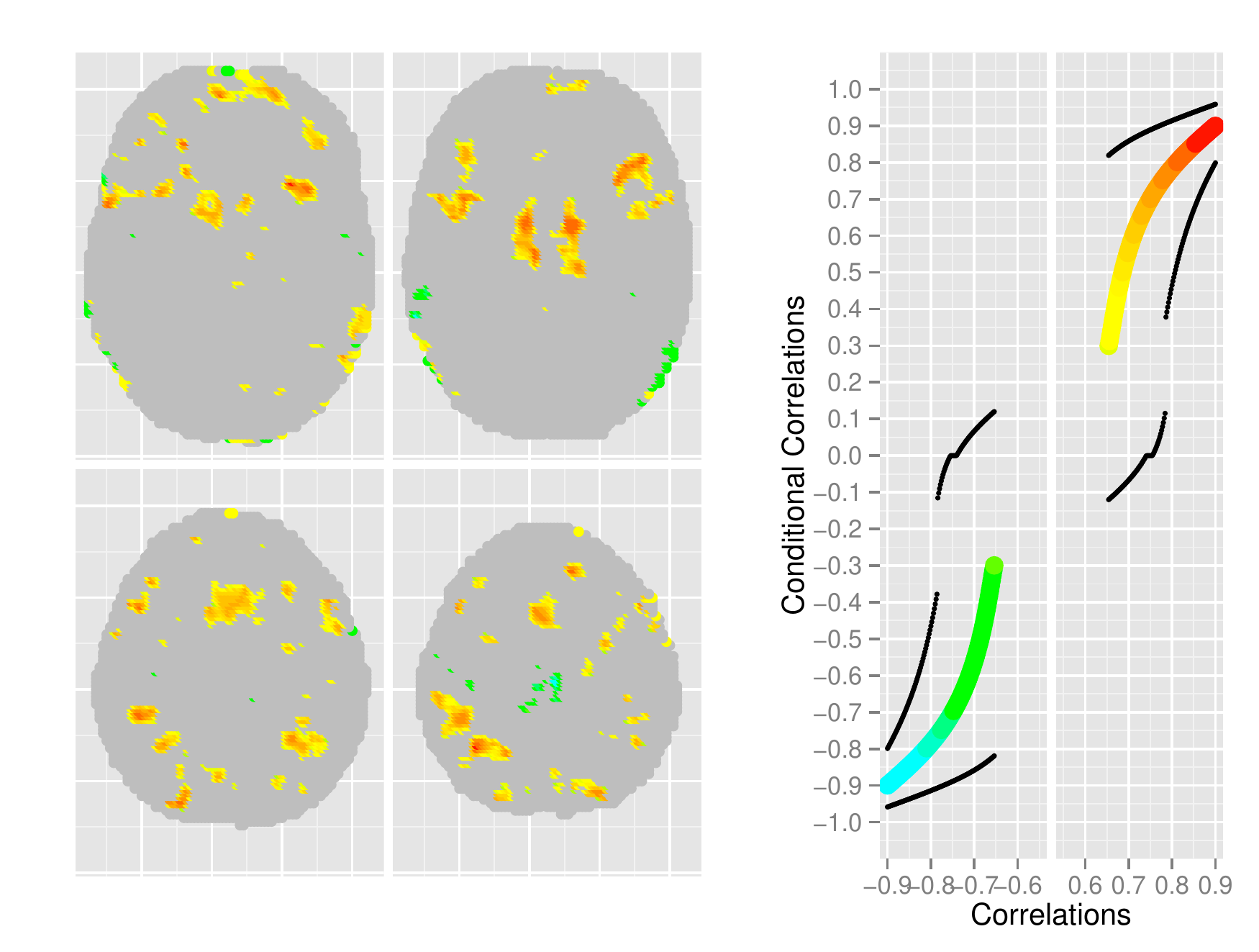} \caption[A Confidence Calibration Plot]{A Confidence Calibration Plot. Observed correlations in significant voxels $FDR\leq 0.1$). The legend is adapted so that it encodes not only the observed value, but also CQC CIs limits and the conditional estimates of the correlations. Confidence intervals are the vertical range from the band above the observed values to the band below.\label{fig:CCPplot}}
\end{figure}

\end{knitrout}

\section{Discussion}\label{sec:conclusion}
This work addresses the challenge of selective estimation in the context of neuroimaging. We proposed a method for accounting for the estimation bias caused by the two step inference procedure in which the observed values used for selection are later reported as point estimates. The proposed method is computationally efficient and is implemented within an accompanying software package. 

We wish to emphasize that despite the fact that the proposed approach significantly reduces the bias compared with the option of reporting the observed values, the estimation bias is still prevalent even when this method is used. It is therefore important to report confidence intervals along with point estimates, as proposed by \cite{rosenblatt2014voodoo}, and our Confidence Calibration plot implements it.


One of the most appealing properties of the proposed method is that it is a compromise between soft thresholding and hard thresholding. When data points are close to the threshold there is much uncertainty about whether it is an upper tail observation of a much smaller correlation or a 
truly high correlation and therefore the estimator is shrunk towards zero. When the observed correlation is high it can be safe to assume that the observation is truly generated by a high underlying true correlation and therefore very little shrinkage is needed. Bayesian considerations may also result in shrunk estimators, although they do not cater directly to the selection process and they require in addition the specification of a prior.

We have demonstrated via simulations that in realistic fMRI settings our method is more accurate and exhibits less bias than the standard approach of reporting the same correlation values used for selection. Additionally, our method does not involve discarding observation and therefore it has more power to detect active voxels compared to the previously proposed data splitting approach. This is especially important since it was made clear that the main goal of social-neuroscience studies is detecting the active brain regions and that the exact value of the correlation is of lesser interest, rendering the data splitting approach as unattractive even if we are willing to pay for unbiasedness with high variance.

\appendix
\newpage
\section{Conditional Estimation After BH Selection}\label{app:BH}
In section \ref{sec:method} we presented the conditional estimation method for a fixed threshold. However, in many cases we are interested in performing selection in such a way as to control quantities such as the FWER by using a Bonferroni procedure or the FDR by using the BH procedure. The issue which arises in the case of selection rules such as BH, is that the threshold is data dependent and not constant. In a simulation study, \cite{rosenblatt2014voodoo} showed that CQC CIs indeed control the FCR when conditioned on the effective threshold determined by the data.

\begin{knitrout}
\definecolor{shadecolor}{rgb}{0.969, 0.969, 0.969}\color{fgcolor}\begin{figure}[ht]

\includegraphics[width=\maxwidth]{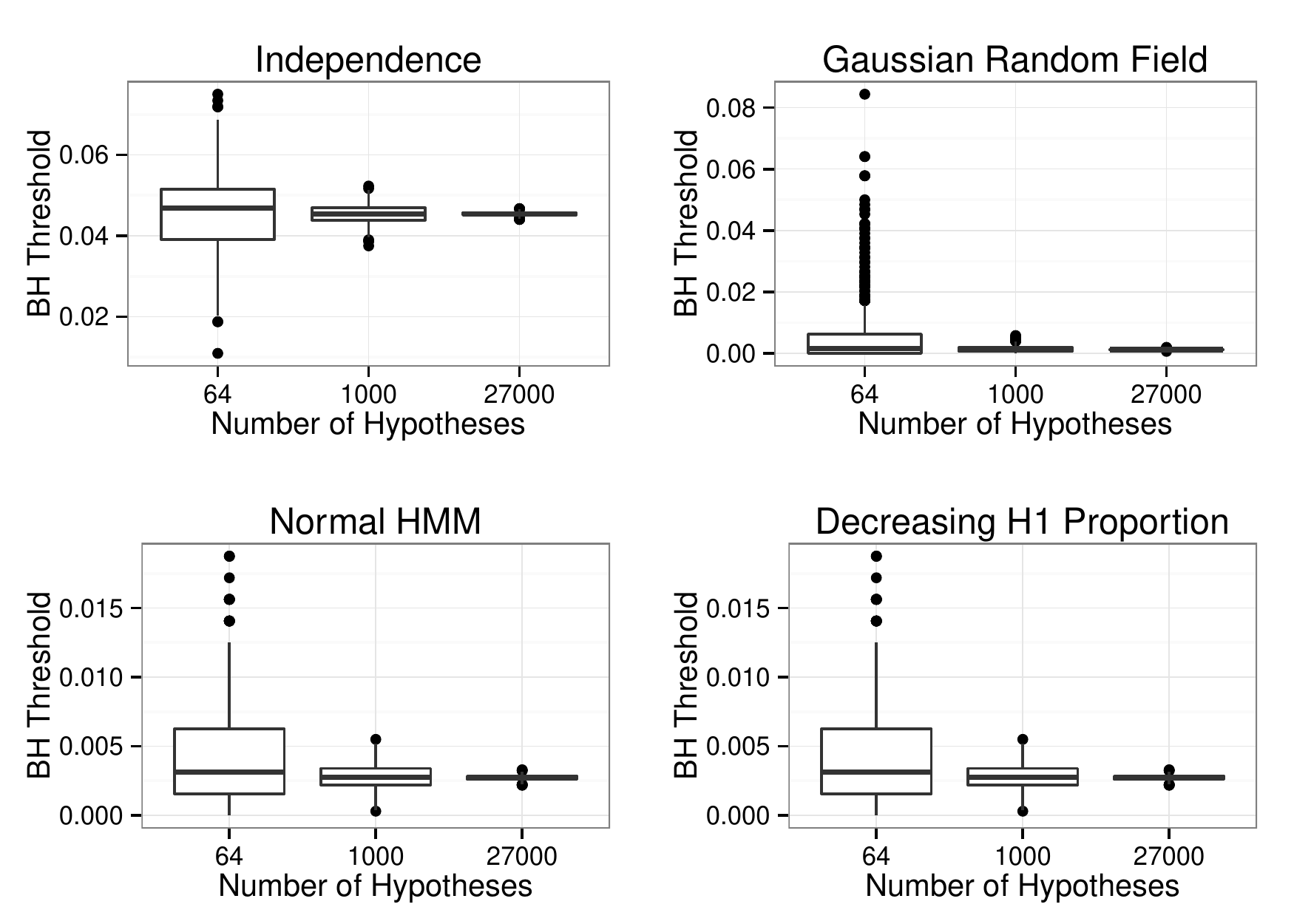} \caption[Demonstration of the convergence of the BH Threshold as the number of hypotheses tested grow under different dependence structures]{Demonstration of the convergence of the BH Threshold as the number of hypotheses tested grow under different dependence structures. In the top left plot the p-values are independent, in the top right plot the p-values are sampled from a Gaussian Random Field, in the the bottom left plot the p-values are sampled from a Normal HMM with three states and in the bottom right plot the p-values are independent but the proportion of alternative hypotheses decreases to zero as the number of hypotheses grow.\label{fig:BHthreshold}}
\end{figure}

\end{knitrout}

Conditioning on the threshold found by the BH procedure can be justified by the following theorem which is theorem 5 of \citet{storey2004strong}. Denote by $m$ the number of hypothesis tested and by $m_i$, $i\in\{0,1\}$ the number of null $p$-values and alternative $p$-values. Furthermore, denote by $V(t)$ the number of false positives and by $S(t)$ the number of true positives as a function of a threshold $t\in(0,1]$.
\begin{theorem}
Assume that (1) $\lim_{m\rightarrow\infty}V(t)/m_0=G_0(t)$, $\lim_{m\rightarrow\infty}S(t)/m_1=G_1(t)$ almost surely for each $t\in(0,1]$, where $G_0$ and $G_1$ are continuous functions; (2) The limiting distribution of the $p$-values under the null hypothesis is stochastically greater than or equal to the uniform distribution, $0<G_0(t)\leq t$ for each $t\in(0,1]$; (3) $\lim_{m\rightarrow\infty}(m_0/m):=\pi_0$ exists. Then the BH threshold converges to a constant threshold $t_{\alpha}$.
\end{theorem}

Condition 1 holds for independent $p$-values, but also under $\emph{weak dependence}$, \citeauthor{storey2004strong} give several examples for dependence structures under which this condition holds- finite blocks, ergodic dependence and certain mixing distributions. This result shows that whenever assumptions 1-3 hold the BH threshold is asymptotically consistent (in the number of hypotheses) for $t_\alpha$ and therefore, the conditional estimator based on the BH threshold is consistent for the estimator based on the "correct" threshold of $t_\alpha$. 

Figure \ref{fig:BHthreshold} demonstrates that the BH threshold indeed converges under different dependence structures. The threshold converges when the noise is generated by a Gaussian Random Field (GRF), when the data is generated from finite state HMM and even when we let the ratio of non-null hypothesis tend to zero.

\newpage
\section{Artificial fMRI Data}\label{app:fmri}
\begin{knitrout}
\definecolor{shadecolor}{rgb}{0.969, 0.969, 0.969}\color{fgcolor}\begin{figure}[t]

\includegraphics[width=\maxwidth]{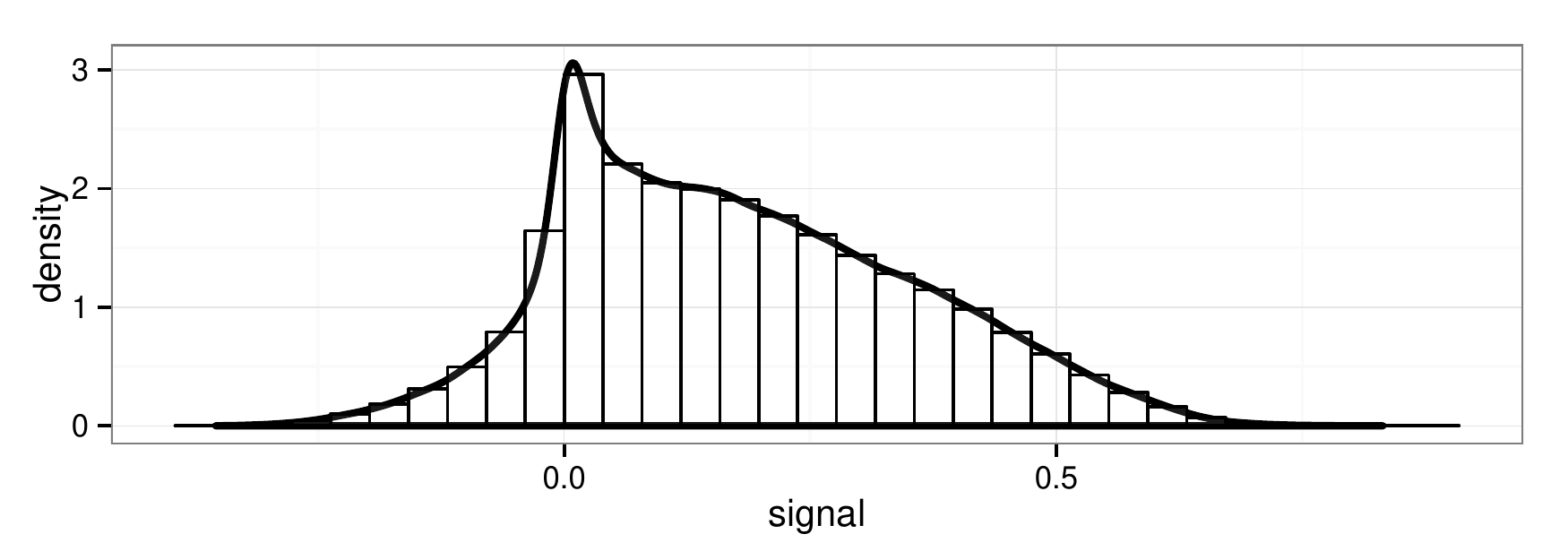} \caption[Denstiy and histogram of the extracted signal]{Denstiy and histogram of the extracted signal\label{fig:signalPlot}}
\end{figure}

\end{knitrout}

\begin{knitrout}
\definecolor{shadecolor}{rgb}{0.969, 0.969, 0.969}\color{fgcolor}\begin{figure}[t]

\includegraphics[width=\maxwidth]{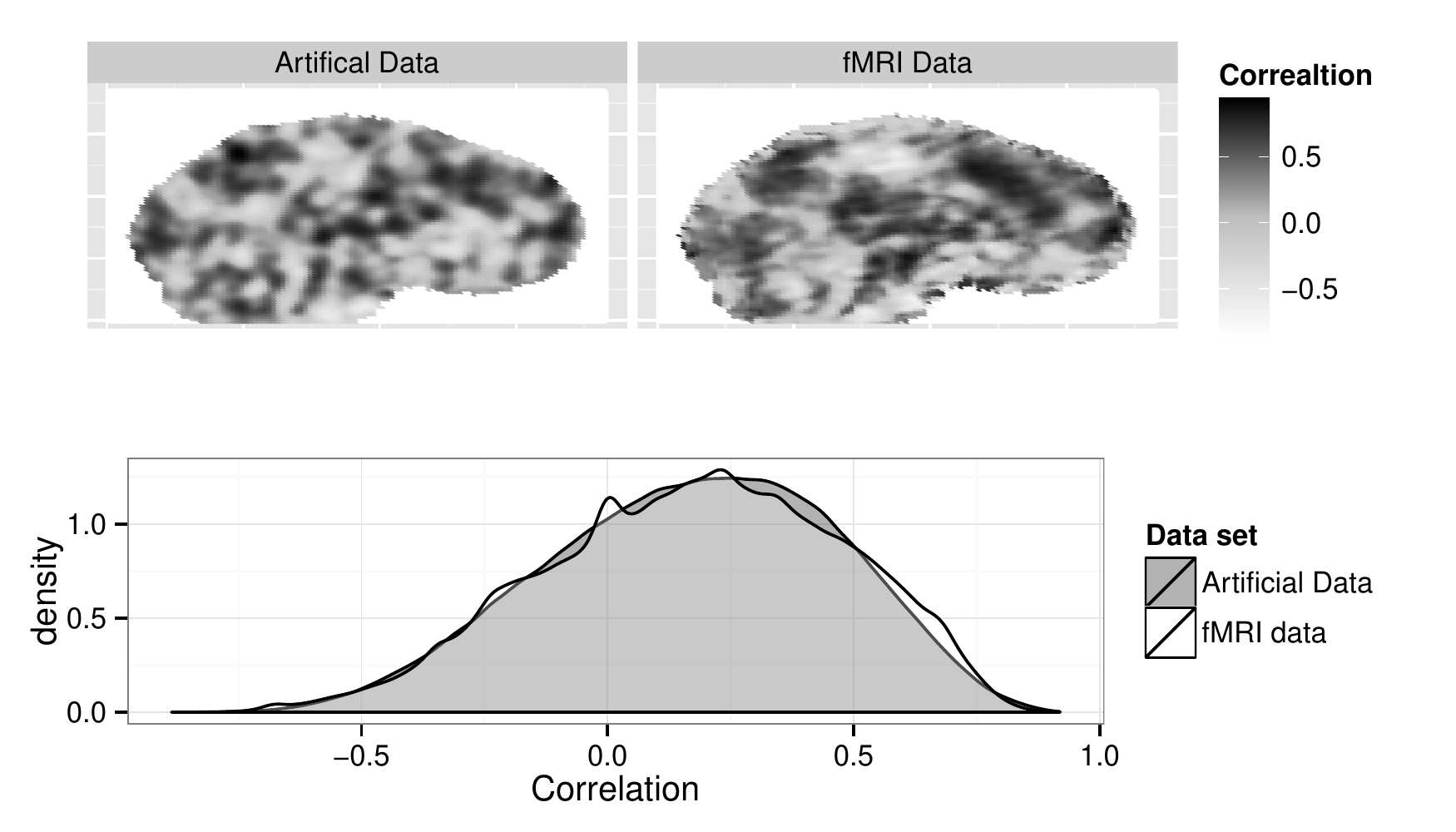} \caption[Artifical and real fMRI data]{Artifical and real fMRI data. In the top plot two fMRI images of brain slices are presented next to one another, on the left the original data and on the right an artifical data. In the bottom plot the marginal density of the artificial data is overlaid on top of the estimated density of the original data.\label{fig:fMRIdemonstration}}
\end{figure}

\end{knitrout}

In this section we describe our method for generating data similar to fMRI data. We assume that after Gaussianization, the observed fMRI data is the sum of a signal, the distribution of which is unknown to us, and a GRF noise with average variance of $(n-3)^{-1}$. In order to generate data that will have similar structure to the real fMRI data we first extract a signal distribution from the raw fMRI data. We do so by fitting a two component Gaussian mixture model to the data where one of the means is set to zero and shrinking the observed correlations towards the means of the Gaussian distributions using a Gaussian kernel with standard deviation $\sigma=0.13$. The extracted signal is then smoothed using a three dimensional Gaussian kernel. The distribution of the extracted signal is presented in Figure \ref{fig:signalPlot}.

We join the extracted signal with a GRF noise to obtain an artificial fMRI data set. The resulting data are presented in Figure \ref{fig:fMRIdemonstration} together with the original data. The artificial data is quite similar in structure to the real data albeit being a bit smoother. 

\newpage
\bibliography{selectiveBib}{}
\bibliographystyle{plainnat}

\end{document}